\documentstyle[prl,multicol,aps]{revtex} 
\input{epsf}
\begin{document}

\wideabs{
\title{Firewalls, Disorder, and Percolation in Epidemics}
\author{C. P. Warren$^1$, L. M. Sander$^1$, and I. M. Sokolov$^2$}
\address{$^1$ Michigan Center for Theoretical Physics,\\
University of Michigan, Ann Arbor, MI, USA, 48109-1120, USA\\
$^2$ Theoretische Polymerphysik,
Unversit\"at Freiburg, Freiburg, Germany}
\date{\today} 
\maketitle 

\begin{abstract}
We consider a spatial model 
related to bond percolation
for the spread of a disease that includes
variation in the susceptibility to infection. We work on a lattice
with random bond strengths and show that with strong disorder,
i.e. a wide range of variation of susceptibility, patchiness in the
spread of the epidemic is very likely, and the criterion for
epidemic outbreak depends strongly on the disorder. These results
are qualitatively different from those of standard models in
epidemiology, but correspond to real effects.
\end{abstract}
\pacs{PACS numbers: 87.23.Cc, 89.75.Hc, 05.40.-a, 64.60.Ak}
}

The most commonly used models for the spread of an epidemic assume
perfect mixing: i.e., all individuals are able to infect all others.
In an inhomogeneous population the susceptibility to infection is
replaced by its average. To account for spatial effects many workers
use reaction-diffusion equations to describe traveling waves of
infection\cite{Murray1}. 
In both cases important phenomena are not
reproduced by the models. For
example, it is common experience that in plant diseases\cite{Jeger2,Bailey3}
islands of
susceptible individuals can be protected by a band of immune ones, a
`firewall'. A map of the infected areas looks patchy. Even in human
diseases such as AIDS these effects seem to be important\cite{Friedman4}. 
Such effects do not occur in perfect mixing or reaction-diffusion
theories in which susceptibilities are replaced by averages.
In this work we give a simple model for these
phenomena using ideas taken from percolation theory. We show that for
inhomogenous populations, the variability of susceptiblity to
infection among individuals gives rise to qualitatively new effects.

Percolation theory\cite{Stauffer5,Grassberger6} 
is a natural way to think about populations
which are not well mixed. Consider a spatially
random distribution
of susceptible individuals which occur with probability $p$.  A (site)
percolation model for an epidemic
is a lattice algorithm where, with probability $p$, a site is
occupied and susceptible sites infect their susceptible neighbors.  
For   diseases where some
agents travel long distances, `small-world' theory\cite{Small7} extends the
percolation model. In this approach a few sites are considered to be
neighbors of distant sites.

Percolation accounts for the existence of firewalls and islands because
near the percolation threshold only part of the lattice belongs to
the spanning cluster.
However, a close look makes such an application suspect. Islands and
firewalls occur only in a rather narrow transition
region $\delta p$ near the critical probability, $p_c$. 
In nature it would be unlikely that a fine-tuning
of parameters to be near $p_c$ would occur. This kind of
difficulty always plagues attempts to apply theories of
critical phenomena in the natural world.  Here, we show that
introducing disorder in the susceptibility to infection solves
this problem in an elegant way because 
{\em it broadens the transition region for the outbreak}, 
so that a generic epidemic
could produce islands and firewalls. 

There is a further unexpected
result of this study. By definition, an epidemic starts 
when a sick individual
infects more than one other before she recovers, i.e. 
$R_o \ge 1$, where $R_o$, the reproduction number, is the
mean number of infections produced by a site.
The usual expression for $R_o$ is 
$xS(0)\tau$ where $x$ is the probability per unit time to infect
neigbors, $S(0)$ the number which can be infected, and $\tau$ the
time to recovery. We will show that disorder can change this formula
by a large amount.

Our model is of the SIR (Susceptible, Infected, Recovered) type. We
use bond percolation: each site of a square lattice is occupied by an
individual which can be infected by neighbors connected by bonds (four
in our case). To
account for variability the probability per unit time of infection
along a given bond is chosen from a distribution $f(x)$. After
exactly  $\tau$
time steps any infected site recovers. We start with one $I$ in the
middle of the lattice and the
rest $S$. We say that we have an epidemic when the $I$ and $R$ sites
span a large lattice, i.e. reach the edges. 
A snapshot is shown in Figure 1. If the recovery
time, $\tau$, is too small the epidemic will die out, and if $\tau$ is
large enough it will persist. A critical value of $\tau$ plays the role of
$p_c$. In Figure 2 we show the spanning probability, $M$, for a
256 x 256 lattice as a function of $\tau$ for two different choices of
$f(x)$. $M$ is the probability to span starting from a single site.
We should note that $M$ is the
same as the mass of the infinite cluster in percolation 
theory\cite{Stauffer5}
i.e., the fraction of sites which belong
to the spanning cluster. This is true 
because the probability to span starting at any given site
is the probability that the
site is a member of the infinite cluster.

We have studied two classes of $f$'s which we call weak and 
strong disorder. An example of weak disorder is:
\begin{equation}
f_w(x) = 1/x_{max}, \quad 0\le x \le x_{max}. 
\label{weak}
\end{equation}
Other functions with a rather narrow range of $x$ give similar
effects, and the results do not differ very much from the 
case with no disorder\cite{Grassberger6}.

Strong disorder refers to broad 
distribution such as:
\begin{equation}
f_s(x) = C/x \quad x_{min} \le x \le 1; \quad C=1/|\ln(x_{min})|
\label{strong}
\end{equation} 
Any strongly skewed distribution should give similar results
to those quoted here. 
Distribution functions of this type were studied\cite{Ambegaokar} 
in random resistor networks, and are
known, in that context, to give different
behavior from the ordered case.  

The black portion of Figure 1 corresponds to sites that are protected
by the weak bonds at the edge of the infected cluster, the firewall. 
Effects of this type occur 
only in the transition region of Figure 2. The transition region
is very broad for strong disorder. Firewalls and large islands
of uninfected sites arise purely from the statistics of the problem,
as in any percolation problem. 

Ordinary bond percolation has some bonds present and some absent, 
which appears to be
quite different from our approach. If $x$ takes on a single value, 
the relationship was pointed out by
Grassberger\cite{Grassberger6}:
the probability to infect a neighbor before recovery 
is $p = 1-(1-x)^{\tau}$. Thus $p$ is the fraction of bonds 
completed in a given epidemic. If we have disorder we write:
\begin{equation}
p = 1 - \int (1-x)^{\tau} f(x) dx.
\label{prob}
\end{equation} 
For a square lattice $p_c = 1/2$, so we expect that when $p$ in Equation 
\ref{prob} is greater than $1/2$ we will have an epidemic. 
We verify this
in the inset to Figure 2 where plot $M$ against $p$ for various
choices of $f(x)$
and compare them to ordinary bond percolation 
on the same lattice. 
 
We can discuss the threshold for outbreak of an epidemic by using 
the usual SIR equations\cite{Murray1} for the perfect mixing case:
\begin{equation}
dS/dt = -xSI, \qquad
dI/dt = xSI - I/\tau
\label{SIR}
\end{equation}
Here there is a rate of recovery $1/\tau$ rather than a fixed recovery
time. However this is qualitatively the same. Clearly,
we have an outbreak if $R_o = xS(0)\tau > 1$. 

If we convert Equation \ref{SIR} into a spatial model by adding diffusion terms
then the criterion for outbreak is of the same form \cite{Murray1},  
where $S(0)$ now refers to the initial number of susceptibles which
can be infected by a single infected site, a number of order unity.
For the Grassberger model (or weak disorder) we have a similar looking result:
we put $p =  1-(1-x)^{\tau_c} = p_c = 1/2$. 
For small $x$, $R_o = x\tau_c = O(1)$. 

Strong disorder is quite different. We can estimate
the integral in Eq.(\ref{prob}) using the  $f_s(x)$ of Eq. 
(\ref{strong}) for $x_{min}<<1$. 
We get
\begin{equation}
\int (1-x)^{\tau} f_s(x) dx \approx [-\Gamma -\ln(\tau x_{min})]/|\ln(x_{min})|
\label{estimate}
\end{equation}
where $\Gamma$ is Euler's constant.
 Using this 
estimate we find
$\tau_c \approx 0.56/\sqrt{x_{min}}$.
To estimate $R_o$ we replace $x$ by its average, $\overline{x}$. 
The naive estimate for the reproduction number at
the threshold for outbreak gives:
\begin{equation}
R_o = \tau_c \overline{x} \propto 1/[\sqrt{x_{min}} |\ln(x_{min})|].
\label{naive}
\end{equation}
This estimate for $R_o$
can be arbitrarily large. The effective $R_o$  for
a disordered lattice is very much smaller than $\tau_c \overline{x}$.

This statement may seem reminiscent of the well-known fact \cite{May}
that for diseases such as AIDS the naive formula for $R_o$
does not work if there are very active agents, because active agents 
have an outsize effect
in spreading the disease. However, the results are, in fact, in opposite
directions. Here, the spatial correlations between strong bonds make
them less effective than they might seem to be. A clump
of very infective bonds does not do as much damage as we might
suspect since the infected sites are sharing the same victims.
In the transition region there are always bottlenecks to the
spread of the infections, and thus random spatial correlations
are always important. 

We can understand the broadening of the transition region due to
disorder in the same way. 
Using the expressions in Eq. (\ref{prob}) and Eq. (\ref{estimate})
we find 
$\delta \tau/\tau_c = |\ln(x_{min})| \delta p$.
We can define the transition region, $\delta p$, 
as the range over which $M$ increases from, say, 0.1 to 0.9;
it is of order 0.05.
We see that $\delta \tau/\tau_c$ is large as
$x_{min} \to 0$.

In epidemics of human diseases some (usually a small
number) of individuals travel long distances.
We use a small-world lattice to account for these contacts by adding 
a small fraction, $\phi$, of `long' bonds connecting randomly chosen
sites\cite{Small7}.
The result is shown in Figure 3. Even a very small $\phi$
breaks up the large uninfected regions. The percolation
threshold is lowered, and below the old $p_c$ there are many small regions
of epidemic connected by long bonds. However, {\it the effects of disorder 
remain.} Long distance contacts are not the same as perfect mixing:
firewalls still occur over a substantial range of parameters.

To understand these results we extend the work of Newman and 
Watts\cite{Newman} to our case. The infected islands of Figure 3 
can be considered to 
be nodes of a random graph whose edges are the long bonds. 
The graph percolates if there are two islands 
per long bond\cite{Alon}; this occurs for some
$p<p_c$. 

We need to count the nodes of the random graph. To do this we use 
the average island size defined as $\overline{n}=\sum s^2 n_s/\sum sn_s$ 
where $n_s$ is the number of clusters of size $s$. This is the correct 
average because it counts the probability that a given site (the end
of a long bond) belongs to a cluster \cite{Stauffer5}.
Scaling theory gives  $ \overline{n} =K|p-p_c|^{-\gamma}$ 
where $K$ is a constant, and $\gamma$ is a critical exponent
equal to $43/18 = 2.39$ for our case. 
The number of islands is $N_i=N/\overline{n}$, where $N$ is the total number
of sites. 
The number of long bonds that connect two 
sites that are parts of islands is $B = N\phi [1-(1-p)^4]^2$. Setting
$2N_i=B$ gives a criterion for percolation on a small world lattice with
one adjustable parameter, $K$:
\begin{equation}
\phi = K (p_c-p)^{\gamma} (1-(1-p)^4)^{-2}
\label{fit}
\end{equation}


In Figure \ref{pc} we show the depression of the percolation
threshold as a function of $\phi$. We exhibit data as a function
of $p$ rather than $\tau$. They are related by Eq. (\ref{prob}).
The threshold is obtained by requiring that an epidemic infects
20\% of the lattice sites. (In this case, with long bonds, the 
notion of spanning loses its meaning).
We also show a fit to the formula in Eq. (\ref{fit}) with
$\gamma$ and $K$ as adjustable parameters. The best fit 
gives $\gamma = 2.40$, remarkably close to the theoretical
value quoted above. 
However, as we see in Figure (\ref{small}) there are still 
islands is surrounded by a firewall, and all of the above
considerations carry over in the strongly disordered case. 
In particular, the size of the transition region in terms of
$\tau$ is expanded \cite{tobe}.     

Our model for percolation with strong disorder illustrates two effects: 
there is a mechanism which produces patches of
uninfected but susceptible individuals without fine tuning
of parameters. And for this case $R_o$ depends on the disorder. 
Whether strong disorder is a valid description of 
nature needs to be determined by
consideration of real epidemics. It is known, for example, that 
variability in susceptibility increases when epidemics recur. Our model
may apply best to that case.

We would like to thank C. Simon and J. Koopman who 
taught us what little we know about epidemics.  Supported by DOE grant 
DEFG-02-95ER-45546.

\begin{figure}
\epsfxsize=3.2in
\epsfbox[71 193 563 674]{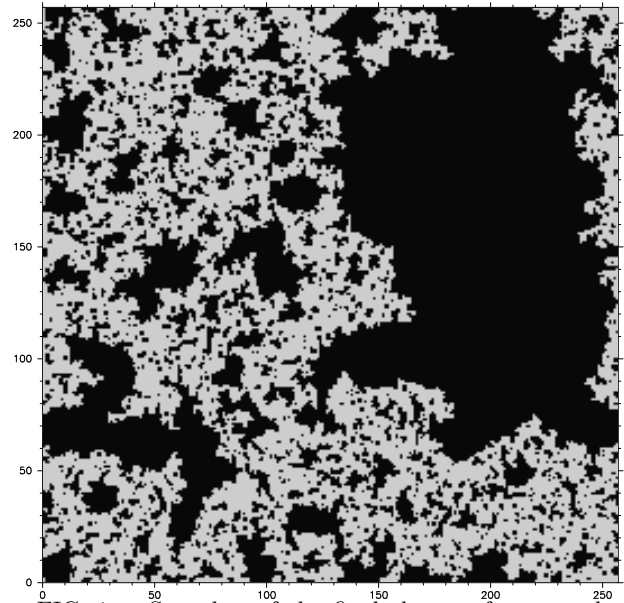}
\caption{ Snapshot of the final cluster of recovered after the epidemic
has died out on a 256x256 lattice with periodic boundary conditions.  
The cluster of recovered is gray, and the uninfected are black.  A strong 
disorder ($x_{min}=e^{-10}$) bond distribution (\ref{strong}) is used.
} 
\end{figure}

\pagebreak

\begin{figure}
\epsfxsize=3.2in
\epsfbox[0 120 612 700]{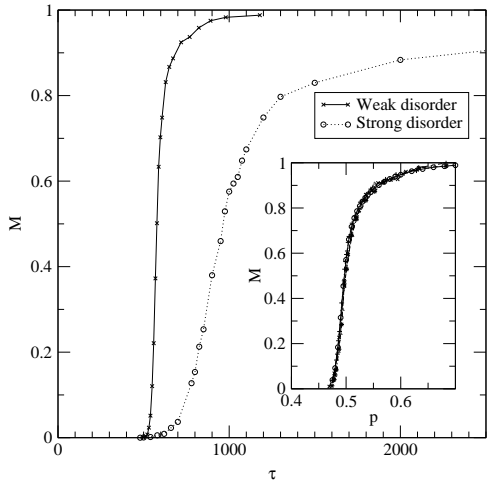}
\caption{The probability, $M$, that an epidemic started from a point
spans the lattice as a function 
of infection time $\tau$ for two different bond distributions.  Strong 
disorder broadens the transition region in $\tau$.
The transition region (e.g. the range $0.1< M <0.9$) is about 12 times
larger for strong disorder than for weak.
For $f_w$ we use
$x_{max}=0.003$, $f_s$ uses $x_{min}=e^{-15}$. 
Averages are over  2000 simulations on a 200x200 
lattice.
[Inset]  Data collapse of 
$M$ as a function of $p$, compared with
bond percolation. The bond distributions used are: (o) bond percolation,
(+) $f_s$ with $x_{min}=e^{-15}$, (x) $f_s$ with 
$x_{min}=e^{-7}$, (*) $f_w$ with $x_{max}=.001$, and (A) $f_w$ with
$x_{max}=.01$.
} 
\end{figure}

\begin{figure}
 \epsfxsize=3.2in
\epsfbox[71 193 563 674]{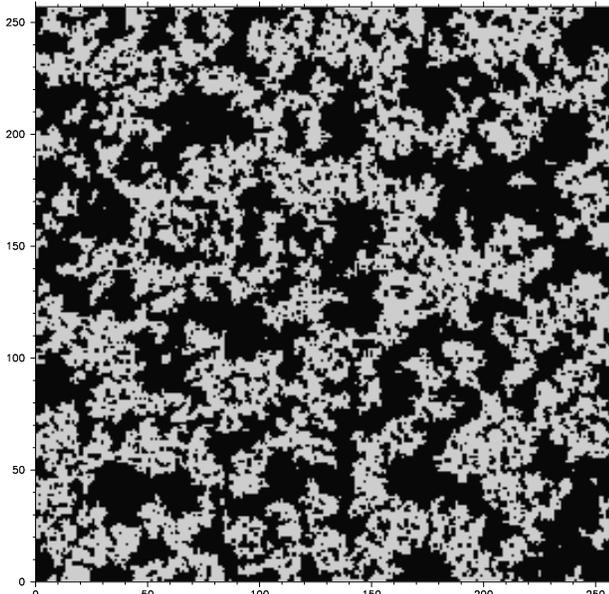}

\caption{Snapshot of the final cluster of recovered on a 256x256 small-world
lattice with periodic boundary conditions and $\phi=0.01$.
}
\label{small}
\end{figure}

\begin{figure}
\epsfxsize=3.4in
\epsfbox{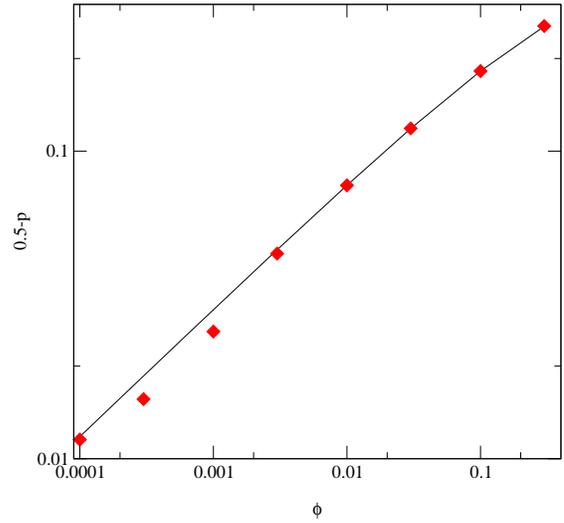}
\caption{Dependence of the percolation threshold on $\phi$ 
($\diamond$)   
compared with the 
scaling theory prediction on a 200x200 lattice.}
\label{pc}
\end{figure}

\end{document}